\documentclass[9pt,onecolumn,twoside]{arxiv}
\usepackage[]{geometry}
\usepackage[scriptsize,tight]{subfigure}
\usepackage{graphicx}
\usepackage{cite}
\usepackage{multirow}
\usepackage{dcolumn}
\usepackage{bm}
\usepackage{color,soul}
\usepackage{hyperref}

\articletype{inv} 

\runningtitle{Interactions between ionized glyphosate and carbon nanotube}

\runningauthor{Silva \textit{et al.}}

\title{Computational study of interactions between ionized glyphosate and carbon nanotube: An alternative for mitigating environmental contamination}

\author[1]{H.~T.~Silva}
\author[1]{L.~C.~S.~Faria}
\author[2]{T.~A.~Aversi-Ferreira}
\author[1,$\ast$]{I.~Camps}
\correspondingauthoraffiliation[$\ast$]{icamps@unifal-mg.edu.br}

\affil[1]{Laborat\'orio de Modelagem Computacional - \emph{La}Model,
Instituto de Ci\^{e}ncias Exatas - ICEx. Universidade Federal de Alfenas -
UNIFAL-MG, Alfenas, Minas Gerais, Brazil}

\affil[2]{Laboratory of Biomathematics, Institute of Science and Tecnology - ICT, Federal University of Alfenas - UNIFAL-MG, Poços de Caldas, Minas Gerais, Brazil}

\begin{abstract}
The extensive use of glyphosate in agriculture has raised environmental concerns due to its adverse effects on plants, animals, microorganisms, and humans. This study investigates the interactions between ionized glyphosate and single-walled carbon nanotubes (CNT) using computational simulations through semi-empirical tight-binding methods (GFN2-xTB) implemented in the xTB software. The analysis focused on different glyphosate ionization states corresponding to various pH levels: G1 ($pH < 2$), G2 ($pH \approx 2-3$), G3 ($pH \approx 4-6$), G4 ($pH \approx 7-10$), and G5 ($pH > 10.6$). Results revealed that glyphosate in G1, G3, G4, and G5 forms exhibited stronger interactions with CNT, demonstrating higher adsorption energies and greater electronic coupling. The neutral state (G2) showed lower affinity, indicating that molecular protonation significantly influences adsorption. Topological analysis and molecular dynamics confirmed the presence of covalent, non-covalent, and partially covalent interactions, while the CNT+G5 system demonstrated moderate interactions suitable for material recycling. These findings suggest that carbon nanotubes, with their extraordinary properties such as nanocapillarity, porosity, and extensive surface area, show promise for environmental monitoring and remediation of glyphosate contamination.
\end{abstract}


\begin{document}
\maketitle
\thispagestyle{firststyle}
\vspace{-13pt}

\section{INTRODUCTION}
\label{Sec:Intro}
Pesticides comprise a group of substances used in agriculture, including insecticides, fungicides, herbicides, rodenticides, molluscicides, and nematicides. Among them, glyphosate (N-phosphonomethyl glycine) stands out as one of the most widely used herbicides in agricultural, forestry, and urban environments worldwide due to its effectiveness in controlling weeds~\cite{Feng_2020,Tudi_2021}. However, residues from this pesticide have been associated with the contamination of terrestrial and aquatic ecosystems, causing serious environmental toxicity.

Among the effects observed on ecosystems are decreased reproduction, loss of biomass, and reduced soil surface activity. In addition, there is a potential risk of human exposure, which can cause epilepsy, act as an endocrine disruptor, damage placental cells, and reduce the enzyme aromatase~\cite{Carvalho_2017,Feng_2020,Gomes_2023}. Thus, efforts have been made to develop technologies capable of detecting, removing, and monitoring the presence of this compound in different environmental compartments~\cite{Espinoza_Montero_2020}. Glyphosate belongs to the chemical group of phosphonate amino acids and has glycine as its precursor, exhibiting an amphoteric and zwitterionic behavior. At neutral $pH$, it can coexist with a positive charge in the amino group and a negative charge in the phosphonate group~\cite{Jayasumana_2014,Mendes_2020}. In the presence of water and depending on the of the medium, glyphosate can exist in different states of ionization. When isolated or in a gaseous state, it has a sum of charges equal to zero; however, in the presence of water at any $pH$ value, it will present different degrees of ionization. Thus, the ionic form of glyphosate directly influences its affinity for adsorbent surfaces, altering the interaction mechanisms involved, such as electrostatic forces, hydrogen bonds, and $\pi-\pi$ interactions. Ionized forms are important for obtaining an accurate and realistic description of molecular interactions in systems designed for the removal of this contaminant, as this impacts the efficiency and selectivity of the materials applied in removal or detection~\cite{Feng_2020}.

Traditional extraction methods, such as biodegradation, photocatalysis, electrochemical processes, membrane separation, oxidation, and adsorption, have not been sufficient in treating these compounds, as they do not promote the total degradation of this substance, which requires post-treatment steps for adsorbent materials or solid wastes, which are complex and economically unfeasible~\cite{Haque_2011,Yang_2018,Gaberell_2019,Espinoza_Montero_2020}. On the other hand, biological processes can generate metabolites such as AMPA, with greater toxicity potential if the operating conditions are not controlled, and a post-treatment with other technologies is recommended to achieve a superior degradation performance~\cite{Feng_2020}. Given these limitations, there is a need to develop more effective, selective, and environmentally safe alternatives for the treatment of this contaminant. Carbon nanotubes stand out for their physicochemical properties, which allow them to efficiently adsorb pesticides or their degradation products, in addition to their use in the production of filters for various pollutants~\cite{Arora_2020,Jampilek_2020}.

Carbon nanotubes are formed by rolling one or more sheets of graphene into a concentric shape, with a diameter in nanometric dimensions and a hollow internal cavity~\cite{Rahman_2019,Aligayev_2022}. The structure of CNTs gives them extraordinary physical and chemical properties, including a large specific surface area and superior thermal, mechanical, and electrical properties. Thus, CNT-based materials have been used to adsorb organic pollutants from wastewater based on their porous structure and large specific surface area~\cite{Peng_2021}.

Considering the limitations of traditional methods and the complexity associated with the glyphosate ion speciation, it is necessary to seek new approaches that take into account the chemical form of the glyphosate molecule and the properties of the adsorbent material~\cite{Feng_2020,Sen_2021}.
In this context, computational modeling is a tool that can anticipate the behavior of real systems and thus predict and describe the dynamics of molecular systems, as they are capable of quantifying interaction energies, mapping electronic distributions, and delineating molecular mechanisms~\cite{Barreiro_1997,Barreiro_2002}. Therefore, this study conducted a computational investigation of the interactions between pure carbon nanotube and the different ionized forms of glyphosate found in nature, which aims to elucidate the influence of $pH$ on affinities, binding mechanisms, and, consequently, the effectiveness and selectivity of removing this agrochemical.

\section{MATERIALS AND METHODS}
\label{Sec:Method}

Single-walled carbon nanotube with chirality (10,0), a zigzag configuration with semiconductor character, and dimensions of 7.83~\AA~in diameter and 12.78~\AA~in length were used.

The analysis of the interactions between glyphosate and carbon nanotube was performed using the semi-empirical tight binding method implemented in the xTB (extended tight binding) software package, which is self-consistent, accurate, and includes electrostatic contributions from multipoles and density-dependent dispersion~\cite{Bannwarth_2020}. For this purpose, the geometry of the molecules was optimized, as this process sought to find the lowest potential energy configuration by adjusting the positions of the atoms. Thus, the atomic coordinates were modified until the calculated energy reached its lowest value and its ground state~\cite{Schlegel_2011}. An extreme level was used to optimize the structures, with a convergence energy of $5\times10^{-8}\,E_h$ and a gradient norm convergence of $5\times10^{-5}\,E_h/a_0$ (where $a_0$ is the Bohr radius)~\cite{Aguiar_2024}.

\begin{figure}[tbph]
\centering
\includegraphics[width=8cm]{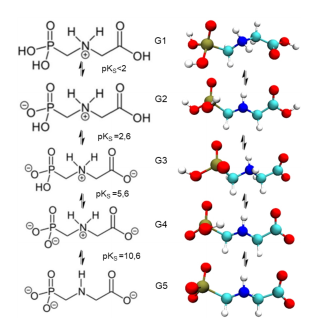}
\caption{\label{Fig:Glypho} 2D and 3D representations of the different ionized forms of glyphosate as a function of pKa values (for interpretation of the references to color in this figure legend, the reader is referred to the web version of this article).}
\end{figure}

The complexes analyzed were formed between a single-walled carbon nanotube (CNT) and glyphosate at different degrees of ionization. The degrees of ionization were defined based on the acid dissociation constants of glyphosate (pKa: 2.0; 2.6; 5.6, and 10.6), that is, the acid dissociation constants of glyphosate, which will be defined by the $pH$ of the medium where the molecule is found, being G1 ($pH < 2$), G2 ($pH \approx 2-3$), G3 ($pH \approx 4-6$), G4 ($pH \approx 7-10$) and G5 ($pH > 10.6$)~\cite{Herath_2019}. The structure of glyphosate for each degree of ionization considered in this work are shown in Figure~\ref{Fig:Glypho}. For organizational purposes, the nomenclature used to identify the simulated systems was defined as CNT+GY, where CNT refers to carbon nanotube and GY represents the specific ionized form of glyphosate considered in each complex.

\begin{figure}[tbph]
\centering
\includegraphics[width=10cm]{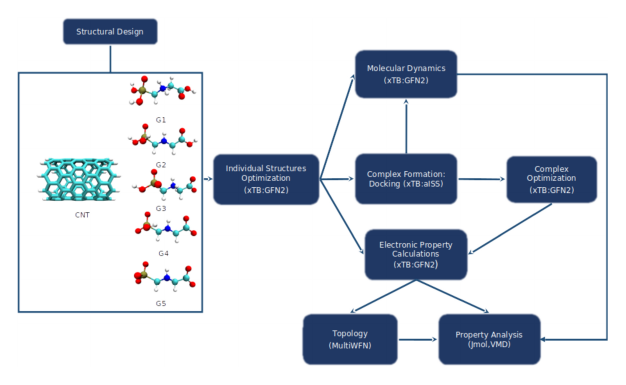}
\caption{\label{Fig:Methods} Flowchart of the computational procedure used for analysis of interactions between carbon nanotube and glyphosate.}
\end{figure}

The calculations were performed according to the sequence of procedures presented in the flowchart in Figure~\ref{Fig:Methods}. Initially, the geometries of the isolated structures of the nanotube and glyphosate were optimized. Next, the coupling process was performed using automated interaction site mapping (aISS). To do this, it was sought to identify regions of accessibility on the surface of the nanotube (molecule A), followed by three-dimensional (3D) screening to identify $\pi-\pi$ interactions in various directions. Subsequently, adjustments were made to identify the most favorable orientations of molecule B (glyphosate) around molecule A (nanotube)~\cite{xTB-dock}.

To classify the generated structures, interaction energy was used, which by default adopts a two-step refinement protocol based on genetic algorithms, in which one hundred structures with the lowest interaction energies are selected to ensure that conformations not detected in the initial screening are included. During this two-step genetic optimization procedure, each glyphosate molecule positions was randomly combined around the carbon nanotube. Then, 50\% of the structures undergo random mutations in both position and angle. After ten interactions of this search process, ten complexes with the lowest interaction energies were selected. Thus, the structure with the lowest interaction energy was then chosen as input for the optimization of the complex~\cite{Aguiar_2024}.

The electronic properties were determined using the spin polarization scheme, and the following parameters for the CNT+GY systems were calculated: highest occupied molecular orbital energy (HOMO, $\varepsilon_H$); the lowest
unoccupied molecular orbital energy (LUMO, $\varepsilon_L$); energy gap between the HOMO and LUMO orbitals ($\Delta \varepsilon = \varepsilon_H - \varepsilon_L$).

To investigate the mobility of charge carriers and evaluate the electronic interactions between nanotube and glyphosate molecules, electronic transfer integrals were calculated using the dimer projection method (DIPRO). In this analyses, $J_{oc}$ denotes the transport of holes (occupied molecular orbitals) and when it has high values, it indicates a greater decoupling between the two fragments, that is, the stronger is the repulsion or stabilization effect between filled orbitals. $J_{un}$ represents the electron transport (unoccupied molecular orbitals) and, when it has high values, indicates a strong interaction between the orbitals, which facilitates electronic transitions, while $J$ represents the total charge transfer, encompassing both hole and electron transport between occupied and unoccupied molecular orbitals, respectively. Higher $J$ values indicate chemical reactivity and changes in the electronic state of the molecule with a strong coupling between the two fragments~\cite{Rauk_2004,Albright_2013,Kohn_2023}.

The adsorption energy ($E_{ads}$), calculated as the difference between the energy of the final CNT+GY system ($E_{CNT+GY}$) and the sum of the energies of the initial isolated CNT ($E_{CNT}$) and glyphosate systems ($E_{GY}$), is given by:

\begin{equation}
\label{Eq:bind}
E_{ads} = E_{CNT+GY} - E_{CNT}- E_{GY}.
\end{equation}

To understand and classify the types and strength of interactions formed between glyph\-o\-sate and the carbon nanotube, a study of topological properties was conducted, using the wave function obtained in the calculation of electronic properties~\cite{Aguiar_2024}. This allowed the identification of bond critical points (BCPs) and the quantification of descriptors such as the electron density ($\rho$), the Laplacian ($\nabla^2 \rho$), the electronic localization function (ELF), and the localized orbital locator (LOL), which were analyzed using the MULTIWFN software, which uses the wave function generated during the calculations of electronic properties~\cite{Lu_2012,Multiwfn2}. To evaluate the strength and type of bond between attractive pairs of atoms, only the critical bonding points of type \textbf{(3,-1)} were analyzed, as these are characterized by a minimum electron density along the bonding path between two nuclei at the interface of the glyphosate molecule and the nanotube. Thus, we obtained parameters based on a physical observable (electron density) that is free of bias, complementing the wave function or the molecular orbital analysis techniques. This avoids assigning physical meaning to a specific set of orbitals, which, although not devoid of physical meaning, analyses based solely on them may lose important details. In addition, electronic density has the advantage of being analyzable both theoretically and experimentally~\cite{Bader1994,Koch_2024,Fedorov_2025}.

In contrast to the geometric optimization, which seeks to identify the lowest energy structure on a potential energy surface, molecular dynamics (MD) simulations allowed us to examine the movement of glyphosate molecules, enabling a more comprehensive understanding of the dynamic behavior of the system~\cite{Martinez_2003}. The molecular dynamics simulations were conducted at $300\,K$ with a production run time of 100~ps, utilizing a time step of 2~fs and a dump step of 50~fs, where the final configuration was recorded in a trajectory file. These calculations employed the GFN-FF force field, which is specifically designed to balance high computational efficiency with the accuracy typically associated with quantum mechanics methods~\cite{xTB_GFN-FF}.

To characterize the spatial distribution of glyphosate molecules, the radial distribution function (RDF) was used:

\begin{equation}
\label{Eq:RDF}
g(\bf{r}) = \frac{n(\bf{r})}{4 \pi \rho \bf{r}^2 \Delta \bf{r}},
\end{equation}
where $n(\bf{r})$ is the mean number of particles in a shell of width $\Delta \bf{r}$ at distance $\bf{r}$, and $\rho$ is the mean particle density.

Statistically, $g(\bf{r})$ describes the probability of finding a glyphosate molecule at a position $\bf{r}$ from the carbon nanotube, normalized by the average density. This approach is useful in heterogeneous systems such as the one analyzed in this study, as it helps to predict how glyphosate organizes itself in relation to the surface of the nanotube~\cite{Hansen_2013}.

\section{RESULTS AND DISCUSSION}
\label{Sec:Results}

\subsection{Structural and electronic properties}
\label{Sec:Struct-Elect}

\begin{figure}[tbph]
\centering
\includegraphics[width=15cm]{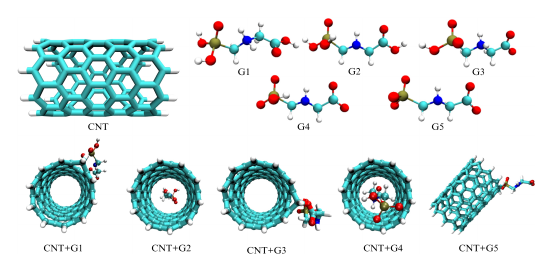}
\caption{\label{Fig:OptStruc} Optimized structures of carbon nanotube, glyphosate, and the complexes formed (for interpretation of the references to color in this figure legend, the reader is referred to the web version of this article).}
\end{figure}

In Figure~\ref{Fig:OptStruc}, the systems are presented after the geometric optimization stage, a fundamental procedure in the investigation of molecular structures and reactivity. This procedure aims to determine the lowest potential energy configuration by interactively adjusting the atomic coordinates until the system reaches its ground state, that is, minimizes the internal forces~\cite{Schlegel_2011}. This resulting conformation serves as the basis for subsequent calculations of electronic and interactive properties, which ensures that further analyses are performed based on representative system conformations.

Some structural changes can be observed between the geometries obtained after the interaction between the CNT and the different ionized forms of glyphosate (G1 to G5). These changes can be attributed to the presence of glyphosate, which, when interacting with the surface of carbon nanotube, can modify their electronic properties and affect the geometry of the system. CNT have a hexagonal structure with $sp^2$ hybridization, which is highly sensitive to surface disturbances. Thus, the presence of ionized glyphosate can impact properties such as electrical conductivity, state density, and the electronic affinity of the system~\cite{Meunier_2016,ribeiro-Appl.Surf.Sci.-426-781-2017}.

Table~\ref{Tab:DataResults} shows the adsorption energy values, frontier orbitals energy ($\varepsilon_H$ and $\varepsilon_L$), energy difference between these orbitals ($\Delta \varepsilon$), and electronic coupling parameters ($J_{oc}$, $J_{un}$, and $J$). These properties were used to understand the stability and reactivity of the systems formed by carbon nanotube and glyphosate molecules with different degrees of ionization.

\begin{table}
\caption{Electronic properties of complexes formed between carbon nanotube and glyphosate in different ionization states$^\dagger$.}
\label{Tab:DataResults}
\begin{center}
\setlength\extrarowheight{-3pt}
\begin{tabular}{lrrrrrrrrrrr}
\hline
System & $E_{ads}$ & $\varepsilon_H$ & $\varepsilon_L$ & $\Delta \varepsilon$ & $J_{oc}$ & $J_{un}$ & $J$ \\
\hline
\hline \\
CNT+G1 & -3.27 & -11.08 & -11.04 & 0.04 & 0.01 & 0.01 & 0.01 \\

CNT+G2 & -1.02 & -8.90 & -8.90 & 0.00 & 0.018 & 0.01 & 0.018 \\

CNT+G3 & -2.44 & -6.65 & -6.90 & 0.05 & 0.02 & --- & --- \\

CNT+G4 & -4.49 & -5.12 & -5.08 & 0.05 & 0.01 & --- & --- \\

CNT+G5 & -9.84 & -3.40 & -3.36 & 0.04 & 0.01 & --- & --- \\
\hline
\end{tabular}
\begin{flushleft}
\tiny {$^\dagger$ Electronic properties were calculated using an electronic temperature of $300\,K$. Subsequently, the HOMO and LUMO energy levels were determined using the Fermi energy. All energies are in units of eV.
}
\end{flushleft}
\end{center}
\end{table}

The adsorption energy values shown in Table~\ref{Tab:DataResults} vary greatly for the complexes as there is a change in the ionized form of glyphosate. This can be correlated with the protonation states of the molecule ($pKa < 2.0; 2.6; 5.6; 10.6$) depending on the $pH$ of the medium~\cite{Jayasumana_2014,Mendes_2020}. In G1 ($pH < 2$), the cationic form of the glyphosate molecule predominates, with a net positive charge, which may favor attractive electrostatic interactions and result in a more negative adsorption energy value. This is consistent with the values of the CNT+G3 and CNT+G5 systems, which are formed by the most protonated forms of glyphosate, favoring interactions via hydrogen bonds, which improves its affinity with the nanotube surface. In CNT+G2, the glyphosate molecule is in its neutral state, with a decrease in its electrostatic attraction capacity, resulting in less negative $E_{ads}$ values~\cite{Feng_2020,Sen_2021}.

\begin{figure}[tbph]
\centering
\includegraphics[width=15cm]{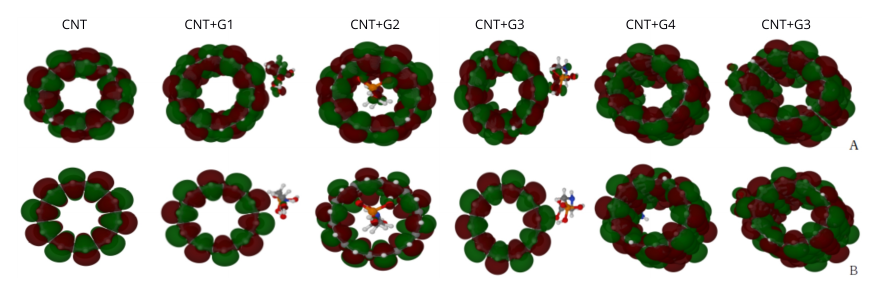}
\caption{\label{Fig:Orbitals} Structure for the LUMO (top panel A) and HOMO (bottom panel B) orbitals of the complexes formed between carbon nanotube and ionized glyphosate (for interpretation of the references to color in this figure legend, the reader is referred to the web version of this article).}
\end{figure}

The energies of the LUMO (Figure~\ref{Fig:Orbitals} top panel, A) and HOMO (Figure~\ref{Fig:Orbitals} bottom panel, B) frontier orbitals in Table~\ref{Tab:DataResults} were used with the other electronic properties to characterize the molecule's ability to donate and accept electrons, respectively. It can be observed that they have different values according to the ionization levels of glyphosate, but with a small variation in gap ($\Delta \varepsilon$). Thus, the system formed by CNT+G2 has the lowest gap value and tends to be the most unstable, while CNT+G4; CNT+G3; CNT+G1 and CNT+G5 have higher GAP values, which indicates greater molecular stability, respectively, that is, they have low reactivity in reactions~\cite{Miessler_2014}. In this context, Reber and Khanna~\cite{Reber_2017} state that electronic distribution can be easily deformed when there are low gaps, resulting in high polarizability.

The feasibility of reusing an adsorbent is correlated with its regeneration capacity. This is an important aspect from an economic and environmental perspective, as it enables a prolonged use of the material, which reduces costs and minimizes environmental impacts~\cite{Zavareh_2018,Krishnamoorthy_2019,Diel_2021}. For systems that obtained very strong interactions, such as CNT+G1, CNT+G4, and CNT+G4, there is difficulty in regenerating and reusing the nanotubes as adsorbents due to the high adsorption energy. Where there were moderate interactions, such as CNT+G5, an easier desorption is possible, which facilitates the recycling and reuse of these materials. For CNT+G2, the adsorption was low, glyphosate may not be efficiently adsorbed, remaining available in the environment.

\subsection{Topological analysis}
\label{Sec:Topo}
By determining the bond critical points (BCPs) in the topological analysis, it was possible to investigate the mechanisms of chemical bond formation and the nature of the intermolecular interactions between glyphosate and carbon nanotube. From the orthodox point of view of the Quantum Theory of Atoms in Molecules (QTAIM), the simultaneous presence of a bonding path and a bond critical point between two atoms is a necessary and sufficient condition for the existence of a chemical interaction between them, although it is widely recognized that alternative interpretations can occur, in the case of weak or very weak interactions~\cite{Van_der_Maelen_2019}.

\begin{figure}[tbph]
\centering
\includegraphics[width=8cm]{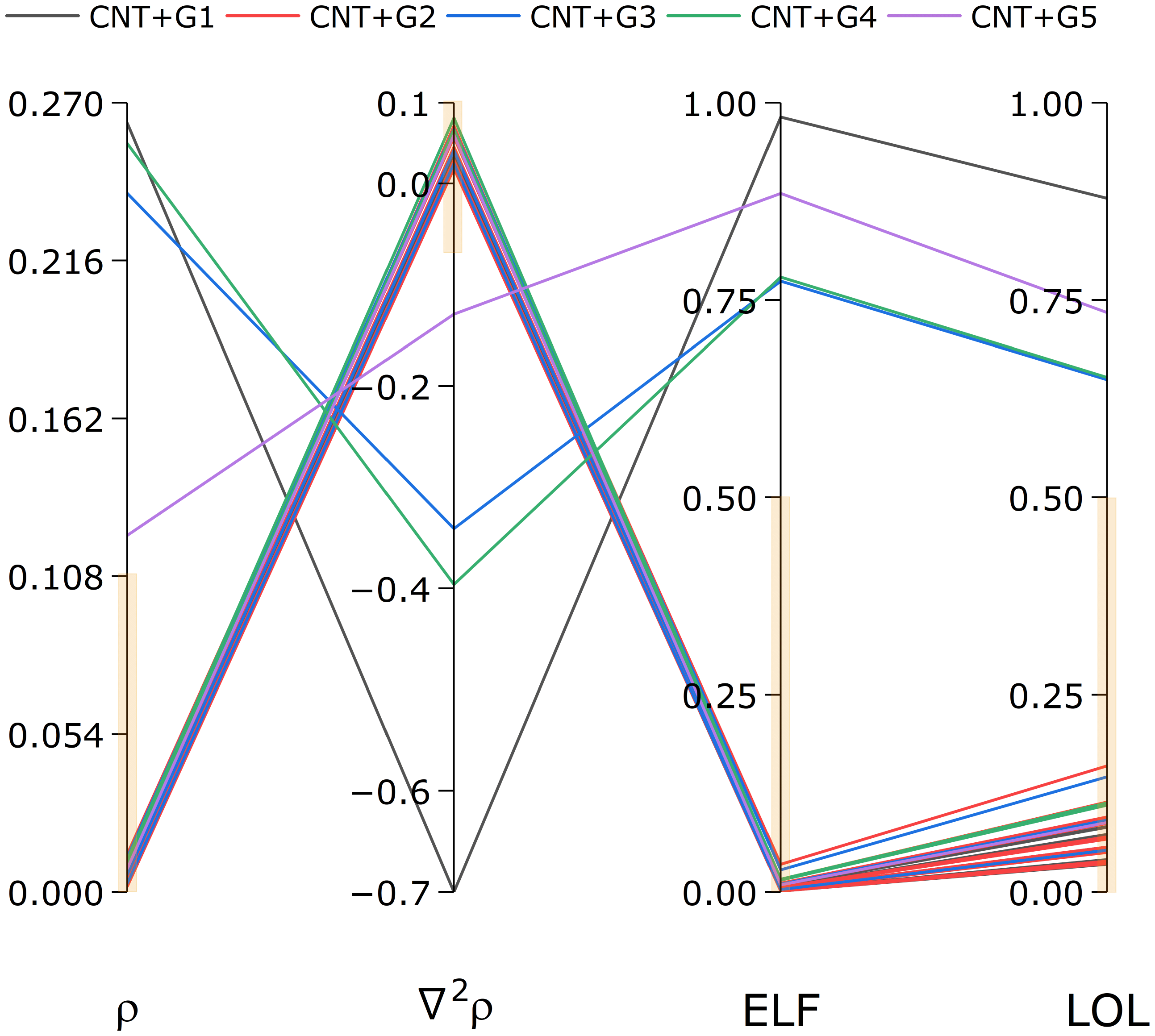}
\caption{\label{Fig:Topo} Topological descriptors, $\rho$, $\nabla^2 \rho$, ELF and LOL, at the bond critical points (BCP) for each complex. In orange, the non-covalent regions for each descriptor. (For interpretation of the references to color in this figure legend, the reader is referred to the web version of this article.)}
\end{figure}

In this context, in order to assess the strength and type of bonding between attractive pairs of atoms, only critical points type \textbf{(3,-1)} (known as bond critical points, BCP) were analyzed, as these are characterized by a minimum electronic density along the bonding path between two nuclei at the interface of the glyphosate molecule and the nanotube.

In our calculations were identified 31 BCP, distributed as follows for the systems: 8 for CNT+G1, 14 for CNT+G2, 4 for CNT+G3, 3 for CNT+G4, and 2 for CNT+G5.  Figure~\ref{Fig:Topo} shows the profiles of electronic density ($\rho$), Laplacian ($\nabla^2 \rho$), electronic localization function (ELF) and orbital locator (LOL) for all the BCP. The bond type (covalent or non-covalent) can be classified based on the electron density ($\rho$) and its Laplacian ($\nabla^2 \rho$) values. Bonds with $\rho > 0.20~a.u.$ and $\nabla^2 \rho < 0$ are characterized as covalent, while those with $\rho < 0.10~a.u.$ and $\nabla^2 \rho > 0$ indicate non-covalent interactions~\cite{Matta2007}.  The
non-covalent regions are marked in orange for each descriptor in Figure~\ref{Fig:Topo}.

\begin{figure}[tbph]
\centering
\includegraphics[width=\textwidth]{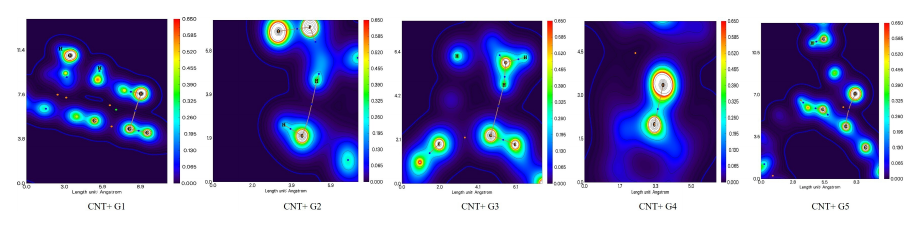}
\caption{\label{Fig:Rho} 2D representation of the electronic density for the complexes formed between the nanotube and ionized glyphosate. (For interpretation of the references to color in this figure legend, the reader is referred to the web version of this article.)}
\end{figure}

The electronic density map (Figure~\ref{Fig:Rho} and Figure~\ref{Fig:Topo}) shows that there are regions where there is electronic accumulation, with high $\rho$ values. This indicates stable interactions, especially for the CNT+G1, CNT+G3 and CNT+G4 systems, and suggests that they are covalent bonds. On the other hand, CNT+G2 and CNT+G5 have lower $\rho$ values or positive Laplacian, which can be seen in the maps as the more dispersed and less defined regions between glyphosate and nanotube, indicating van der Waals bonds.

\begin{figure}[tbph]
\centering
\includegraphics[width=\textwidth]{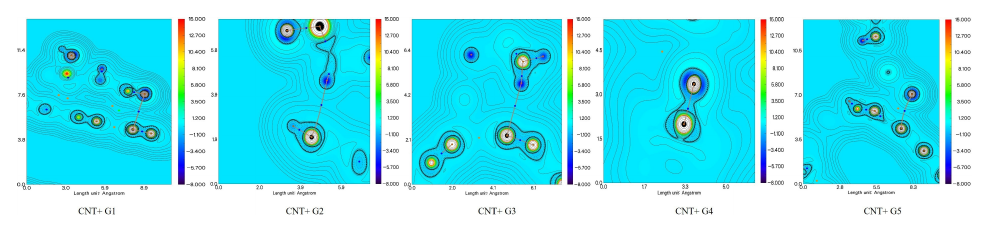}
\caption{\label{Fig:Lap} 2D representation of the electron density Laplacian for CNT–Glyphosate complexes. (For interpretation of the references to color in this figure legend, the reader is referred to the web version of this article.)}
\end{figure}

For the Laplacian (Figure~\ref{Fig:Lap}), negative values are observed at only one critical point for CNT+G1, CNT+G3, CNT+G4, and CNT+G5, which is a strong indication of covalent bonds with concentrated density. For the other points, positive values of $\nabla^2 \rho$ were found, which is typical of non-bonding interaction zones such as van der Waals and electronic depletion regions~\cite{Cabeza_2009,Van_der_Maelen_2019}.

\begin{figure}[tbph]
\centering
\includegraphics[width=\textwidth]{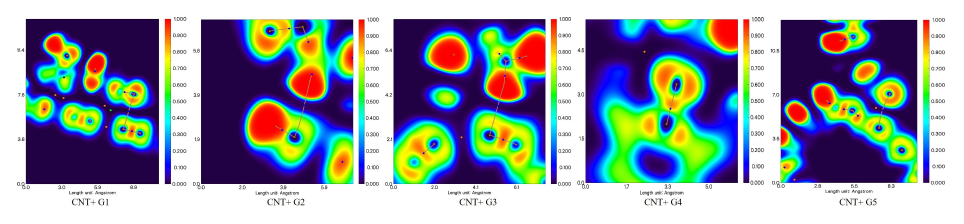}
\caption{\label{Fig:ELF} 2D representation of the electron localization function (ELF) for CNT–Glyphosate complexes. (For interpretation of the references to color in this figure legend, the reader is referred to the web version of this article.)}
\end{figure}

The ELF function map for the molecular plane is shown in Figure~\ref{Fig:ELF}. Its values range from 0 to 1 and were used to identify the domains where electrons can be located, as it defines the regions where there is a high probability of finding electron pairs. In addition, these values can be used to determine the nature of the interactions, since the higher the ELF in the BCPs, the more shared the bond is~\cite{Koumpouras_2020,Khartabil_2025}.

For critical points where the Laplacian was negative, values close to 1 were obtained, indicating an ideal location for these electrons, which reinforces the presence of covalent character at these points. In addition, with the changes in the glyphosate ionization, there are regions with intermediate ELF values. This indicates that there may be partially covalent bonds at these critical points, corroborating those systems such as CNT+G3 and CNT+G5, in which the anionic forms of glyphosate are present, exhibiting electrostatic hypersensitivity, favoring adsorption~\cite{Michalski_2019}.

\begin{figure}[tbph]
\centering
\includegraphics[width=\textwidth]{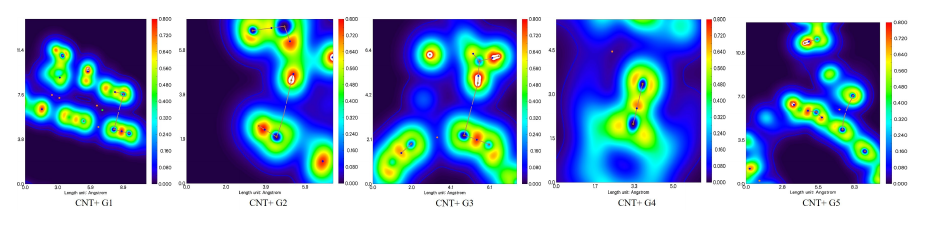}
\caption{\label{Fig:LOL} 2D representation of the localized orbital locator (LOL) for CNT–Glyphosate complexes. (For interpretation of the references to color in this figure legend, the reader is referred to the web version of this article.)}
\end{figure}

Complementarily, the LOL maps presented in Figure~\ref{Fig:LOL} reinforce the analyses presented for the other parameters described, as there are regions of electron pair density accumulation, corroborating the zones identified by ELF and Laplacian, which shows more densely organized interactions in the CNT+G1, CNT+G3, and CNT+G4 systems, confirming the covalent nature of the interactions. For the CNT+G5 system, the LOL values were lower than 0.088, suggesting less orbital organization and non-covalent interactions, despite the relatively high electron density at one of the points.

Thus, the topological analyses based on electronic density that were presented were complementary to the analysis techniques using wave functions or molecular orbitals, being based on a physical observable that is bias-free, as it avoids assigning physical meaning to a specific set of orbitals. Although orbitals are not devoid of physical meaning, analyses based solely on them may lose important details. In addition, electronic density has the advantage of being analyzable both theoretically and experimentally~\cite{Bader1994,Nguyen_2020,Fedorov_2025} .

\subsection{Molecular Dynamics Simulations}
\label{MolDyn}

\begin{figure}[tbph]
\centering
\includegraphics[width=15cm]{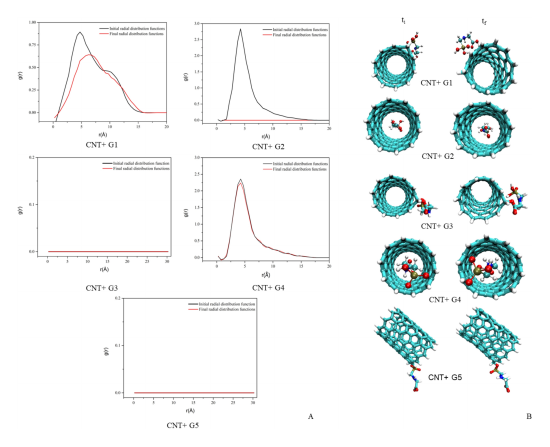}
\caption{\label{Fig:MolDyn} (A) Initial and final radial distribution functions for the complexes calculated with VMD. (B) Initial and final configurations for the complexes ($t_i = 0\,ps$, $t_f = 100\,ps$).}
\end{figure}

Molecular dynamics (MD) simulations were performed on the energetically optimized geometries of all complexes. The results revealed distinct binding behaviors, confirming what was observed in the analysis of electronic and topological properties.
The spatial distribution of molecular interactions was analyzed using radial distribution functions (RDFs) for systems containing glyphosate at its respective degree of ionization interacting with the nanotube. This allowed us to verify the relative probability of finding the glyphosate molecule at a distance r from the CNT surface. Figure~\ref{Fig:MolDyn} presents the RDF analyses and illustrates the initial configurations calculated using VMD software.

To quantify the spatial organization, all RDF profiles were deconvoluted using Gaussian fitting. The positions of the resulting peaks represent the most likely locations for glyphosate molecules in each system. In CNT+G1, an initial peak was observed between 4~7~\AA, while in the final curve there was an attenuation of the peak, suggesting that a rearrangement had occurred. These characteristics indicate weak interactions (electrostatic or hydrophobic), which is consistent with the results obtained for the adsorption energy of this system. The CNT+G3 and CNT+G4 systems maintained stable bonding interactions, with covalent or partially covalent character. This characteristic is corroborated by the high electronic density, ELF, and LOL, in addition to a negative Laplacian at least at one bond critical point, suggesting a potential attraction of the nanotube to the glyphosate with different degrees of chemical speciation. For CNT+G4, there is a slight shift of the peak in the final RDF to the left, indicating that glyphosate approached the nanotube, suggesting a more intense interaction than the initial one. Although no similar behavior is observed for CNT+G3, there is a strong confinement of the molecule inside the nanotube, which can be attributed to an intense and localized interaction.

In the CNT+G2 system, there is significant mobility of glyphosate around the nanotube structure without establishing a preferential interaction or a stable adsorption site, which suggests predominantly weak interactions and an absence of localized adsorption, as analyzed in the topological analysis. The CNT+G5 system showed no evidence of significant interaction in the topological data and RDF analysis, indicating an absence of adsorption. Thus, the results suggest that when glyphosate is in the ionized form G1, G3, and G4, it has the ability to interact with nanotube, while speciation in the form G2 and G5 does not interact stably with the CNT surface~\cite{Cheng_2005}.

\section{CONCLUSIONS}
\label{Sec:Conc}

The results obtained in this study demonstrate the significant impact of glyphosate ionization on the electronic, energetic, and interaction properties of the complexes analyzed. The geometric optimization of each molecule resulted in structures with minimum total energy and verified structural changes, which were related to changes in the ionization of glyphosate molecules.

As for the analysis of adsorption energies, it was possible to observe that the stability of the complexes varies according to the degree of ionization of glyphosate, with the CNT+G1, CNT+G3, and CNT+G4 complexes showing stronger interactions, which may increase the efficiency in adsorption processes but reduces the possibility of regenerating the adsorbent with the reuse of the system, while CNT+G2 had weak interactions. On the other hand, moderate interactions were observed for the CNT+G5 system, suggesting balance and stability, making it viable from both an environmental and economic standpoint, as it allows for the reuse of nanotubes.

The analysis of electronic properties (HOMO, LUMO, energy gap), it was possible to confirm that the ionization of glyphosate molecules influences the reactivity and stability of the complexes. Our results shown that the electronic stability of the systems depends on the chemical form of glyphosate in the medium, that is, the charge of the molecule acts as a modulator of the chemical interactions between glyphosate and nanotubes.

Through the topological analysis, it was possible to understand the nature of the interactions between ionized glyphosate and carbon nanotube. The CNT+G1, CNT+G3, and CNT+G4 systems presented characteristics of covalent bonds, while the other systems had non-covalent bonds. Molecular dynamics (MD) simulations provided a temporal perspective of the interactions, making it possible to observe the stability and behavior of the complexes over time. The systems with greater electronic affinity maintained stable interactions during the simulation, while others showed relative mobility or less persistence at the nanotube interface.

In conclusion, carbon nanotubes show promising in the development of materials for the detection and capture of glyphosate molecules regardless from the form in which they are ionized in the environment, and can be applied in environmental monitoring and remediation. Thus, in future studies, it is recommended to study the influence of functionalization with carboxyl and hydroxyl groups on the interaction properties between carbon nanotubes and glyphosate, in addition to analyzing the economic feasibility and practical application for the large-scale production of this nanomaterial.

\section*{Acknowledgements}
\section*{CRediT authorship contribution statement}
\textbf{H. T. Silva}: Formal analysis, Investigation, Writing-original draft, Writing-review \& editing.
\\
\textbf{L. C. S. Faria}: Formal analysis, Investigation, Writing-original draft, Writing-review \& editing.
\\
\textbf{T. A. Aversi-Ferreira}: Formal analysis, Investigation, Writing-original draft, Writing-review \& editing.
\\
\textbf{I. Camps}: Conceptualization, Formal analysis, Methodology, Project administration, Resources, Software, Supervision, Writing-review \& editing.

\section*{Declaration of competing interest}

The authors declare that they have no known competing financial interests or personal relationships that could have appeared to influence the work reported in this paper.

\section*{Data availability}
\label{data_avail}
The raw data required to reproduce these findings are available to download from Zenodo repository~\cite{Heberson_zenodo.16994309}.

\section*{Acknowledgements}
I.C. is grateful to the Brazilian funding agency CNPq for the research scholarship (304937\allowbreak/2023-1). This study was financed in part by the Coordena\c{c}\~ao de Aperfei\c{c}oamento de Pessoal de N\'{\i}vel Superior-Brasil (CAPES)-Finance code 001. Part of the results presented here were developed with the help of CENAPAD-SP (Centro Nacional de Processamento de Alto Desempenho em S\~ao Paulo) grant UNICAMP/FINEP-MCT, and the National Laboratory for Scientific Computing (LNCC/MCTI, Brazil) for providing HPC resources of the Santos Dumont supercomputer and CENAPAD-UFC (Centro Nacional de Processamento de Alto Desempenho, at Universidade Federal do Cear\'a).

\newpage

\begin{thebibliography}{10}
\expandafter\ifx\csname url\endcsname\relax
  \def\url#1{\texttt{#1}}\fi
\expandafter\ifx\csname urlprefix\endcsname\relax\def\urlprefix{URL }\fi
\expandafter\ifx\csname href\endcsname\relax
  \def\href#1#2{#2} \def\path#1{#1}\fi

\bibitem{Feng_2020}
D.~Feng, A.~Soric, O.~Boutin, Treatment technologies and degradation pathways
  of glyphosate: A critical review, Sci. Total Environ. 742 (2020) 140559
  (2020).
\newblock \href {https://doi.org/10.1016/j.scitotenv.2020.140559}
  {\path{doi:10.1016/j.scitotenv.2020.140559}}.

\bibitem{Tudi_2021}
M.~Tudi, H.~Daniel~Ruan, L.~Wang, J.~Lyu, R.~Sadler, D.~Connell, C.~Chu, D.~T.
  Phung, Agriculture development, pesticide application and its impact on the
  environment, Int. J. Environ. Res. Public. Health 18 (2021) 1112 (2021).
\newblock \href {https://doi.org/10.3390/ijerph18031112}
  {\path{doi:10.3390/ijerph18031112}}.

\bibitem{Carvalho_2017}
F.~P. Carvalho, Pesticides, environment, and food safety, Food Energy Secur. 6
  (2017) 48--60 (2017).
\newblock \href {https://doi.org/10.1002/fes3.108}
  {\path{doi:10.1002/fes3.108}}.

\bibitem{Gomes_2023}
L.~K.~V. Gomes, L.~H.~V. Gomes, J.~C.~M. Amaral, P.~P. Vidal, I.~T. d.~S.
  Gomes, A.~M. d.~J. Chaves~Neto, A.~F.~G. Neto, Molecular dynamics of carbon
  nanotube with fipronil and glyphosate pesticides, The Journal of Engineering
  and Exact Sciences 9 (2023) 16128--01e (2023).
\newblock \href {https://doi.org/10.18540/jcecvl9iss6pp16128-01e}
  {\path{doi:10.18540/jcecvl9iss6pp16128-01e}}.

\bibitem{Espinoza_Montero_2020}
P.~J. Espinoza-Montero, C.~Vega-Verduga, P.~Alulema-Pullupaxi, L.~Fernández,
  J.~L. Paz, Technologies employed in the treatment of water contaminated with
  glyphosate: {A} review, Molecules 25 (2020) 5550 (2020).
\newblock \href {https://doi.org/10.3390/molecules25235550}
  {\path{doi:10.3390/molecules25235550}}.

\bibitem{Jayasumana_2014}
C.~Jayasumana, S.~Gunatilake, P.~Senanayake, Glyphosate, hard water and
  nephrotoxic metals: are they the culprits behind the epidemic of chronic
  kidney disease of unknown etiology in {Sri Lanka}?, Int. J. Environ. Res.
  Public. Health 11 (2014) 2125--2147 (2014).
\newblock \href {https://doi.org/10.3390/ijerph110202125}
  {\path{doi:10.3390/ijerph110202125}}.

\bibitem{Mendes_2020}
K.~F. Mendes, R.~N. de~Sousa, A.~F.~S. Laube,
  \href{https://doi.org/10.5772/intechopen.91872}{Current approaches to
  pesticide use and glyphosate-resistant weeds in {Brazilian} agriculture}, in:
  J.~Moudrý, K.~F. Mendes, J.~Bernas, R.~da~Silva~Teixeira, R.~N. de~Sousa
  (Eds.), Multifunctionality and Impacts of Organic and Conventional
  Agriculture, IntechOpen, Rijeka, 2020, Ch.~1 (2020).
\newblock \href {https://doi.org/10.5772/intechopen.91872}
  {\path{doi:10.5772/intechopen.91872}}.
\newline\urlprefix\url{https://doi.org/10.5772/intechopen.91872}

\bibitem{Haque_2011}
E.~Haque, J.~W. Jun, S.~H. Jhung, Adsorptive removal of methyl orange and
  methylene blue from aqueous solution with a metal-organic framework material,
  iron terephthalate ({MOF}-235), J. Hazard. Mater. 185 (2011) 507--511 (2011).
\newblock \href {https://doi.org/10.1016/j.jhazmat.2010.09.035}
  {\path{doi:10.1016/j.jhazmat.2010.09.035}}.

\bibitem{Yang_2018}
Q.~Yang, J.~Wang, X.~Chen, W.~Yang, H.~Pei, N.~Hu, Z.~Li, Y.~Suo, T.~Li,
  J.~Wang, The simultaneous detection and removal of organophosphorus
  pesticides by a novel {Zr-MOF} based smart adsorbent, J. Mater. Chem.A 6
  (2018) 2184--2192 (2018).
\newblock \href {https://doi.org/10.1039/C7TA08399H}
  {\path{doi:10.1039/C7TA08399H}}.

\bibitem{Gaberell_2019}
L.~Gaberell, C.~Hoinkes, Highly hazardous profits. how {Syngenta} makes
  billions by selling toxic pesticides, Tech. rep., PUBLIC EYE (2019).

\bibitem{Arora_2020}
B.~Arora, P.~Attri, Carbon nanotubes ({CNTs): A} potential nanomaterial for
  water purification, J. Compos. Sci. 4 (2020) 135 (2020).
\newblock \href {https://doi.org/10.3390/jcs4030135}
  {\path{doi:10.3390/jcs4030135}}.

\bibitem{Jampilek_2020}
J.~Jamp\'{\i}lek, K.~Kr\'alov\'a, Carbon nanomaterials for agri-food and
  environmental applications, Elsevier, 2020, Ch. 17 - Potential of nanoscale
  carbon-based materials for remediation of pesticide-contaminated environment,
  pp. 359--399 (2020).
\newblock \href {https://doi.org/10.1016/B978-0-12-819786-8.00017-7}
  {\path{doi:10.1016/B978-0-12-819786-8.00017-7}}.

\bibitem{Rahman_2019}
G.~Rahman, Z.~Najaf, A.~Mehmood, S.~Bilal, A.~Shah, S.~Mian, G.~Ali, An
  overview of the recent progress in the synthesis and applications of carbon
  nanotubes, C 5 (2019) 3 (2019).
\newblock \href {https://doi.org/10.3390/c5010003}
  {\path{doi:10.3390/c5010003}}.

\bibitem{Aligayev_2022}
A.~Aligayev, F.~Raziq, U.~Jabbarli, N.~Rzayev, L.~Qiao,
  \href{https://www.sciencedirect.com/science/article/pii/B9780323854573000190}{Chapter
  17 - {Morphology} and topography of nanotubes}, in: Y.~Al-Douri (Ed.),
  Graphene, Nanotubes and Quantum Dots-Based Nanotechnology, Woodhead
  Publishing Series in Electronic and Optical Materials, Woodhead Publishing,
  2022, pp. 355--420 (2022).
\newblock \href
  {https://doi.org/https://doi.org/10.1016/B978-0-323-85457-3.00019-0}
  {\path{doi:https://doi.org/10.1016/B978-0-323-85457-3.00019-0}}.
\newline\urlprefix\url{https://www.sciencedirect.com/science/article/pii/B9780323854573000190}

\bibitem{Peng_2021}
J.~Peng, Y.~He, C.~Zhou, S.~Su, B.~Lai, The carbon nanotubes-based materials
  and their applications for organic pollutant removal: A critical review,
  Chinese Chem. Lett. 32 (2021) 1626--1636 (2021).
\newblock \href {https://doi.org/10.1016/j.cclet.2020.10.026}
  {\path{doi:10.1016/j.cclet.2020.10.026}}.

\bibitem{Sen_2021}
K.~Sen, S.~Chattoraj, Intelligent environmental data monitoring for pollution
  management, Elsevier, 2021, Ch. A comprehensive review of glyphosate
  adsorption with factors influencing mechanism: Kinetics, isotherms,
  thermodynamics study, pp. 93--125 (2021).
\newblock \href {https://doi.org/10.1016/B978-0-12-819671-7.00005-1}
  {\path{doi:10.1016/B978-0-12-819671-7.00005-1}}.

\bibitem{Barreiro_1997}
E.~J. Barreiro, C.~R. Rodrigues, M.~G.~a. Albuquerque, C.~M. R.~d. Sant’Anna,
  R.~B.~d. Alencastro, Molecular modeling: a tool for rational drug design in
  medicinal chemistry, Qu\'{\i}mica Nova 20 (1997) 300--310 (1997).
\newblock \href {https://doi.org/10.1590/S0100-40421997000300011}
  {\path{doi:10.1590/S0100-40421997000300011}}.

\bibitem{Barreiro_2002}
E.~J. Barreiro, C.~A.~M. Fraga, A.~L.~P. Miranda, C.~R. Rodrigues, Medicinal
  chemistry of {N}-acylhydrazones: novel lead-compounds of analgesic,
  antiinflammatory and antithrombotic drugs, Qu\'{\i}mica Nova 25 (2002)
  129--148 (2002).
\newblock \href {https://doi.org/10.1590/S0100-40422002000100022}
  {\path{doi:10.1590/S0100-40422002000100022}}.

\bibitem{Bannwarth_2020}
C.~Bannwarth, E.~Caldeweyher, S.~Ehlert, A.~Hansen, P.~Pracht, J.~Seibert,
  S.~Spicher, S.~Grimme, Extended tight‐binding quantum chemistry methods,
  WIREs Comput. Mol. Sci. 11 (2020) e1493 (2020).
\newblock \href {https://doi.org/10.1002/wcms.1493}
  {\path{doi:10.1002/wcms.1493}}.

\bibitem{Schlegel_2011}
H.~B. Schlegel, Geometry optimization, WIREs Comput. Mol. Sci. 1 (2011)
  790--809 (2011).
\newblock \href {https://doi.org/10.1002/wcms.34} {\path{doi:10.1002/wcms.34}}.

\bibitem{Aguiar_2024}
C.~Aguiar, I.~Camps, Exploring the potential of boron-nitride nanobelts in
  environmental applications: Greenhouse gases capture, Surfaces and Interfaces
  52 (2024) 104874 (2024).
\newblock \href {https://doi.org/10.1016/j.surfin.2024.104874}
  {\path{doi:10.1016/j.surfin.2024.104874}}.

\bibitem{Herath_2019}
G.~A.~D. Herath, L.~S. Poh, W.~J. Ng, Statistical optimization of glyphosate
  adsorption by biochar and activated carbon with response surface methodology,
  Chemosphere 227 (2019) 533--540 (2019).
\newblock \href {https://doi.org/10.1016/j.chemosphere.2019.04.078}
  {\path{doi:10.1016/j.chemosphere.2019.04.078}}.

\bibitem{xTB-dock}
C.~Plett, S.~Grimme, Automated and efficient generation of general molecular
  aggregate structures, Angew. Chem. Int. Ed. 62 (2022).
\newblock \href {https://doi.org/10.1002/anie.202214477}
  {\path{doi:10.1002/anie.202214477}}.

\bibitem{Rauk_2004}
A.~Rauk, Orbital interaction theory of organic chemistry, 2nd Edition, Wiley,
  2000 (2000).
\newblock \href {https://doi.org/10.1002/0471220418}
  {\path{doi:10.1002/0471220418}}.

\bibitem{Albright_2013}
T.~A. Albright, J.~K. Burdett, M.-H. Whangbo, Orbital interactions in
  chemistry, 2nd Edition, Wiley, 2013 (2013).
\newblock \href {https://doi.org/10.1002/0471220418}
  {\path{doi:10.1002/0471220418}}.

\bibitem{Kohn_2023}
J.~T. Kohn, N.~Gildemeister, S.~Grimme, D.~Fazzi, A.~Hansen, Efficient
  calculation of electronic coupling integrals with the dimer projection method
  via a density matrix tight-binding potential, J. Chem. Phys. 159 (2023)
  144106 (2023).
\newblock \href {https://doi.org/10.1063/5.0167484}
  {\path{doi:10.1063/5.0167484}}.

\bibitem{Lu_2012}
T.~Lu, F.~Chen, Multiwfn: A multifunctional wavefunction analyzer, J. Comput.
  Chem. 33 (2012) 580--592 (2012).
\newblock \href {https://doi.org/10.1002/jcc.22885}
  {\path{doi:10.1002/jcc.22885}}.

\bibitem{Multiwfn2}
T.~Lu, A comprehensive electron wavefunction analysis toolbox for chemists,
  {Multiwfn}, J. Chem. Phys. 161 (2024) 082503 (2024).
\newblock \href {https://doi.org/10.1063/5.0216272}
  {\path{doi:10.1063/5.0216272}}.

\bibitem{Bader1994}
R.~F.~W. Bader, Atoms in molecules: a quantum theory, International series of
  monographs on chemistry, Clarendon Press, Oxford, 1994 (1994).

\bibitem{Koch_2024}
D.~Koch, M.~Pavanello, X.~Shao, M.~Ihara, P.~W. Ayers, C.~F. Matta, S.~Jenkins,
  S.~Manzhos, The analysis of electron densities: from basics to emergent
  applications, Chem. Rev. 124 (2024) 12661--12737 (2024).
\newblock \href {https://doi.org/10.1021/acs.chemrev.4c00297}
  {\path{doi:10.1021/acs.chemrev.4c00297}}.

\bibitem{Fedorov_2025}
I.~Fedorov, Topological analysis of electron density in graphene/benzene and
  graphene/{hBN}, Materials 18 (2025) 1790 (2025).
\newblock \href {https://doi.org/10.3390/ma18081790}
  {\path{doi:10.3390/ma18081790}}.

\bibitem{Martinez_2003}
J.~M. Mart\'{\i}nez, L.~Mart\'{\i}nez, Packing optimization for automated
  generation of complex system’s initial configurations for molecular
  dynamics and docking, J. Comput. Chem. 24 (2003) 819--825 (2003).
\newblock \href {https://doi.org/10.1002/jcc.10216}
  {\path{doi:10.1002/jcc.10216}}.

\bibitem{xTB_GFN-FF}
S.~Spicher, S.~Grimme, Robust atomistic modeling of materials, organometallic,
  and biochemical systems, Angewandte Chemie International Edition 59 (2020)
  15665--15673 (2020).
\newblock \href {https://doi.org/10.1002/anie.202004239}
  {\path{doi:10.1002/anie.202004239}}.

\bibitem{Hansen_2013}
J.-P. Hansen, I.~R. McDonald, Theory of simple liquids: withapplications to
  soft matter, Academic Press, 2013 (2013).

\bibitem{Meunier_2016}
V.~Meunier, A.~G. Souza~Filho, E.~B. Barros, M.~S. Dresselhaus, Physical
  properties of low-dimensional $sp^2$-based carbon nanostructures, Rev. Mod.
  Phys. 88 (2016) 025005 (2016).
\newblock \href {https://doi.org/10.1103/RevModPhys.88.025005}
  {\path{doi:10.1103/RevModPhys.88.025005}}.

\bibitem{ribeiro-Appl.Surf.Sci.-426-781-2017}
M.~S. Ribeiro, A.~L. Pascoini, W.~G. Knupp, I.~Camps, {Effects of surface
  functionalization on the electronic and structural properties of carbon
  nanotubes: A computational approach}, Appl. Surf. Sci. 426 (2017) 781--787
  (2017).
\newblock \href {https://doi.org/10.1016/j.apsusc.2017.07.162}
  {\path{doi:10.1016/j.apsusc.2017.07.162}}.

\bibitem{Miessler_2014}
G.~L. Miessler, P.~J. Fischer, D.~A. Tarr, Inorganic chemistry, 5th Edition,
  Pearson, 2014 (2014).

\bibitem{Reber_2017}
A.~C. Reber, S.~N. Khanna, Superatoms: electronic and geometric effects on
  reactivity, Accounts Chem. Res. 50 (2017) 255--263 (2017).
\newblock \href {https://doi.org/10.1021/acs.accounts.6b00464}
  {\path{doi:10.1021/acs.accounts.6b00464}}.

\bibitem{Zavareh_2018}
S.~Zavareh, Z.~Farrokhzad, F.~Darvishi, Modification of zeolite {4A} for use as
  an adsorbent for glyphosate and as an antibacterial agent for water,
  Ecotoxicology and Environmental Safety 155 (2018) 1--8 (2018).
\newblock \href {https://doi.org/10.1016/j.ecoenv.2018.02.043}
  {\path{doi:10.1016/j.ecoenv.2018.02.043}}.

\bibitem{Krishnamoorthy_2019}
R.~Krishnamoorthy, B.~Govindan, F.~Banat, V.~Sagadevan, M.~Purushothaman, P.~L.
  Show, Date pits activated carbon for divalent lead ions removal, J. Biosci.
  Bioeng. 128 (2019) 88--97 (2019).
\newblock \href {https://doi.org/10.1016/j.jbiosc.2018.12.011}
  {\path{doi:10.1016/j.jbiosc.2018.12.011}}.

\bibitem{Diel_2021}
J.~C. Diel, D.~S.~P. Franco, I.~d.~S. Nunes, H.~A. Pereira, K.~S. Moreira,
  T.~A. de~L.~Burgo, E.~L. Foletto, G.~L. Dotto, Carbon nanotubes impregnated
  with metallic nanoparticles and their application as an adsorbent for the
  glyphosate removal in an aqueous matrix, J. Environ. Chem. Eng. 9 (2021)
  105178 (2021).
\newblock \href {https://doi.org/10.1016/j.jece.2021.105178}
  {\path{doi:10.1016/j.jece.2021.105178}}.

\bibitem{Van_der_Maelen_2019}
J.~F. Van~der Maelen, Topological analysis of the electron density in the
  carbonyl complexes {M(CO)\textsubscript{8}(M = Ca, Sr, Ba)}, Organometallics
  39 (2019) 132--141 (2019).
\newblock \href {https://doi.org/10.1021/acs.organomet.9b00699}
  {\path{doi:10.1021/acs.organomet.9b00699}}.

\bibitem{Matta2007}
C.~F. Matta, R.~J. Boyd, {The Quantum Theory of Atoms in Molecules: From Solid
  State to DNA and Drug Design}, John Wiley \& Sons, 2007 (2007).

\bibitem{Cabeza_2009}
J.~A. Cabeza, J.~F. Van~der Maelen, S.~García-Granda, Topological analysis of
  the electron density in the {N}-heterocyclic carbene triruthenium cluster
  {[$Ru_3(\mu-H)_2(\mu_3-MeImCH)(CO)_9]$ ($Me_2Im$ =
  1,3-dimethylimidazol-2-ylidene)}, Organometallics 28 (2009) 3666--3672
  (2009).
\newblock \href {https://doi.org/10.1021/om9000617}
  {\path{doi:10.1021/om9000617}}.

\bibitem{Koumpouras_2020}
K.~Koumpouras, J.~A. Larsson, Distinguishing between chemical bonding and
  physical binding using electron localization function ({ELF}), J. Phys.
  Condens. Matter 32 (2020) 315502 (2020).
\newblock \href {https://doi.org/10.1088/1361-648X/ab7fd8}
  {\path{doi:10.1088/1361-648X/ab7fd8}}.

\bibitem{Khartabil_2025}
H.~Khartabil, A.~Rajamani, C.~Lefebvre, J.~Pilm\'e, E.~H\'enon, A 30‐year
  journey towards an accelerated scheme for visualizing {ELF} basins in
  molecules, J. Comput. Chem. 46 (2025) e70146 (2025).
\newblock \href {https://doi.org/10.1002/jcc.70146}
  {\path{doi:10.1002/jcc.70146}}.

\bibitem{Michalski_2019}
M.~Michalski, A.~J. Gordon, S.~Berski, Topological analysis of the electron
  localisation function ({ELF}) applied to the electronic structure of
  oxaziridine: the nature of {N-O} bond, Struct. Chem. 30 (2019) 2181--2189
  (2019).
\newblock \href {https://doi.org/10.1007/s11224-019-01407-9}
  {\path{doi:10.1007/s11224-019-01407-9}}.

\bibitem{Nguyen_2020}
D.~D. Nguyen, Z.~Cang, G.-W. Wei, A review of mathematical representations of
  biomolecular data, Phys. Chem. Chem. Phys. 22 (2020) 4343--4367 (2020).
\newblock \href {https://doi.org/10.1039/C9CP06554G}
  {\path{doi:10.1039/C9CP06554G}}.

\bibitem{Cheng_2005}
H.~Cheng, A.~C. Cooper, G.~P. Pez, M.~K. Kostov, P.~Piotrowski, S.~J. Stuart,
  Molecular dynamics simulations on the effects of diameter and chirality on
  hydrogen adsorption in single walled carbon nanotubes, J. Phys. Chem. B 109
  (2005) 3780--3786 (2005).
\newblock \href {https://doi.org/10.1021/jp045358m}
  {\path{doi:10.1021/jp045358m}}.

\bibitem{Heberson_zenodo.16994309}
H.~T. Silva, L.~C.~S. Faria, T.~A. Aversi-Ferreira, I.~Camps, {(DATASET+VIDEOS)
  Computational} study of interactions between ionized glyphosate and carbon
  nanotube: An alternative for mitigating environmental contamination.
\newblock \href {https://doi.org/10.5281/zenodo.16994309}
  {\path{doi:10.5281/zenodo.16994309}}.

\end{thebibliography}

\end{document}